\documentclass{aastex631}
\usepackage{bm}
\usepackage{amssymb}
\usepackage{hyperref}
\usepackage{amsmath}
\usepackage{booktabs}
\usepackage{lipsum}
\usepackage{mathtools}
\usepackage[T1]{fontenc} 

\newcommand{\diff}{{\rm d}}

\newcommand{\hankel}[2]{{\cal H}\left[#1\right]\left(#2\right)}
\newcommand{\comment}[1]{}

\shorttitle{FFT for extended source microlensing}
\shortauthors{Sunao Sugiyama}

\graphicspath{{./}{figures/}}

\begin{document}

\reportnum{IPMU22-0007}
\title{FFT based evaluation of microlensing magnification with extended source}

\author[0000-0003-1153-6735]{Sunao Sugiyama}
\affiliation{Kavli Institute for the Physics and Mathematics of the Universe (WPI), UTIAS \\The University of Tokyo, Kashiwa, Chiba 277-8583, Japan}
\affiliation{Department of Physics, The University of Tokyo, 7-3-1 Hongo, Bunkyo-ku, Tokyo 113-0033 Japan}

\email{sunao.sugiyama@ipmu.jp}

\begin{abstract}
The extended source effect on microlensing magnification is non-negligible and must be taken into account for in an analysis of microlensing. However, the evaluation of the extended source magnification is numerically expensive because it includes the two-dimensional integral over source profile. Various studies have developed methods to reduce this integral down to the one-dimensional-integral or integral-free form, which adopt some approximations or depend on the exact form of the source profile, e.g. disk, linear/quadratic limb-darkening profile. In this paper, we develop a new method to evaluate the extended source magnification based on fast Fourier transformation (FFT), which does not adopt any approximations and is applicable to any source profiles. Our implementation of the FFT based method enables the fast evaluation of the extended source magnification as fast as $\sim1$ msec (CPU time on a laptop) and guarantees an accuracy better than 0.3\%. The FFT based method can be used for the template fitting to a huge data set of light curves from the existing and upcoming surveys.
\end{abstract}


\section{Introduction} \label{sec:intro} 
Microlensing is a gravitational lensing phenomenon, where angular separation between multiple images is too small to resolve, though the magnification of flux as a sum of these images can be detected as a function of time \citep{Einstein.Einstein.1916,Refsdal.Bondi.1964,Paczynski.Paczynski.1986}. This effect is caused by the gravitational force, and suitable for searching for faint or dark compact objects, e.g. dwarf stars, planets, and (primordial) black holes. There are various existing and ongoing surveys designed for the microlensing search: the Optical Gravitational Lensing Experiment\footnote{\url{https://ogle.astrouw.edu.pl}} \citep[OGLE;][]{Udalski.Szymanski.2015}, Microlensing Observations Astrophysics \footnote{\url{http://www2.phys.canterbury.ac.nz/moa/}} \citep[MACHO;][]{Alcock.Sutherland.1993}, Massive Astrophysical Compact Halo Objects\footnote{\url{https://wwwmacho.anu.edu.au}} \citep[MOA;][]{Bond.Yock.2001}, Expérience pour la Recherche d’Objets Sombres\footnote{\url{http://eros.in2p3.fr}} \citep[EROS;][]{Aubourg.Gry.1993}, Disk Unseen Objects \citep[DUO;][]{Alard.Bertin.1995}, Wise Microlensing Survey \citep{Shvartzvald.Maoz.2012}, and Korea Microlensing Telescope Network\footnote{\url{https://kmtnet.kasi.re.kr/kmtnet-eng/}} \citep[KMTNet][]{Kim.Kim.2010}. There are studies using other equipment to search for microlensing: primordial black hole studies by \cite{Griest.Lehner.2014} with the public data from NASA Kepler satellite\footnote{\url{https://www.nasa.gov/mission_pages/kepler/main/index.html}} and by \cite{Niikura.Chiba.2019} with Subaru Hyper Suprime-Cam\footnote{\url{https://www.subarutelescope.org/Observing/Instruments/HSC/}}. These surveys provide a huge data set and enable high detectability of the microlensing events, while requiring rapid template fittings to the obtained light curves. In the early era of the microlensing surveys, microlensing magnification templates for point source was widely used for fitting. Templates for extended source is becoming standard in the community of the microlensing surveys these days, because the extended source effect causes non negligible effect on microlensing light curves and because it gives us chance to break the degeneracy between lensing parameters enabling tighter constraint on physical parameters, e.g. lens mass. In many cases of the microlensing analyses, disk profile is used and enough at the stage of microlensing detection, but more general profiles, like linear limb-darkening profile or quadratic limb-darkening profile, are also used in some surveys or in single event analysis after detection \citep{Yoo.Zebrun.2004}.

The evaluation of the extended source magnification is, however, very expensive because it is given by a convolution of the source profile and the point source magnification in two-dimensional plane. \cite{Gould.Gould.1994} approximated this convolution based on the asymptotic behaviour of the point source magnification at closest lens-source approach to factorize out the extended source effect. \cite{Witt.Mao.1994} derived an analytic expression of the extended source magnification with disk profile. The analytic expression is accurate but includes the third kind of elliptic integral which has a singularity point and requires careful evaluation around it. \cite{Yoo.Zebrun.2004} generalized the method of \cite{Gould.Gould.1994} to the linear limb-darkening profile. These methods are convolution-free and hence allow very fast evaluation, while these include the finite systematic error depending on the extended source parameter due to the approximation. \cite{Lee.Bender.2009} proposed a method to perform the two-dimensional convolution using integration variables on a polar coordinate centering at the lens object to avoid the singularity in the point source magnification. Their method can reduce the two-dimensional integral to a one-dimensional integral for disk profile, while it still includes the two-dimensional integral for more general profile like limb-darkening profile and its evaluation is computationally expensive. \cite{Witt.Atrio-Barandela.2019} developed a method to evaluate the extended source magnification by the Taylor series with coefficients depending on the source profile. This method can be generalized to any analytically-given source profile, and achieves 3\% accuracy with up to the third order of the Taylor series. These studies provide sufficient accuracy with reasonable evaluation time given the size of data set from current surveys.

In the coming decade, we will expect percent level photometric accuracy with upcoming surveys, like Euclid satellite mission \footnote{\url{https://sci.esa.int/web/euclid} \citep{Laureijs.Valenziano.2012}}, Vera C. Rubin Observatory’s Legacy Survey of Space and Time \footnote{\url{https://www.lsst.org}} \citep[LSST;][]{Ivezic.Zhan.2019} and the Nancy Grace Roman Space Telescope \footnote{\url{https://roman.gsfc.nasa.gov}} \citep[former WFIRST]{Spergel.Zhao.2015}. These upcoming surveys also expect to find a number of microlensing events detected, $\sim 50,000$ microlensing events by the Nancy Grace Roman Space Telescope for example \citep{Gaudi.Street.2019}. Hence, processing the data sets from these upcoming surveys requires the more accurate and faster evaluation method of the extended source magnification.

In this paper, we introduce a method to evaluate the extended source magnification with fast Fourier transformation (FFT). The formalism of this method is exact and does not adopt any approximations. Even in the numerical evaluation, we can achieve sub-percent level precision as fast as or faster than methods by other studies. While we show our method only for widely-used source profiles, our method can be easily applied to any other source profiles, whether it is analytic or numeric profile. 

The structure of this paper is as follows. In Section~\ref{sec:basics-of-microlensing}, we review the basics of microlensing and introduce the definition of extended source magnification. In Section~\ref{sec:fft-based-method}, we describe the formalism of our method to evaluate the extended source magnification based on FFT. In Section~\ref{sec:result}, we show how the FFT based method works and compare our result to those of other studies. Throughout of this paper, we assume lens object is point like.

\section{Basics of microlensing}\label{sec:basics-of-microlensing}
The magnification for a point source is given by \cite{Paczynski.Paczynski.1986} as 
\begin{align}
    A_{\rm p}(u) = \frac{u^2+2}{u\sqrt{u^2+4}}. \label{eq:point-source-magnification}
\end{align}
Here $u=u(t)$ is the separation of source and lens objects at time $t$ in the unit of Einstein angle in the sky plane, and given as
\begin{align}
    u(t) = \sqrt{u_{\rm min}^2 + \frac{(t-t_0)^2}{t_{\rm E}^2}},
\end{align}
where $u_{\rm min}$ and $t_0$ are the lens-source separation and the time at the closest approach, and $t_{\rm E}$ is the Einstein time characterizing the time scale of a microlensing event.

The extended source magnification with a source profile, $s(u)$, is given by a convolution of the source profile and the point source magnification in  Eq.~\eqref{eq:point-source-magnification} in two-dimensional sky plane,
\begin{align}
    A_{\rm ext}(u) = \int\diff\bm{x}A_{\rm p}(\bm{x})s(\bm{x}+\bm{u}), \label{eq:extended-source-magnification}
\end{align}
where we introduced two-dimensional vectors, $\bm{x}$ and $\bm{u}$ (following normalization condition $u=|\bm{u}|$), and a notation $f(\bm{x})=f(x)$ if the function has rotational symmetry, like $A_{\rm p}(u)$. We assume in this paper the source profile has rotational symmetry, $s(u)=s(\bm{u})$. The source profile is normalized so that $\int\diff\bm{x}s(\bm{x})=1$. Note that while the right-hand side of Eq.~\eqref{eq:extended-source-magnification} includes angular dependence of vector $\bm{u}$, the extended source magnification on the left-hand side is independent of the angular and hence has rotational symmetry, which can be verified in the FFT based formalism described in Section~\ref{sec:fft-based-method}.

Before going to the FFT based formalism, we introduced several source profiles which are widely used in the microlensing or transit experiments. The most simple profile is the disk profile \citep{Gould.Gould.1994,Witt.Mao.1994}: 
\begin{align}
    s_{\rm disk}(u) = \frac{1}{\pi\rho^2} \Theta(u-\rho), \label{eq:disk-profile}
\end{align}
where $\Theta(x)$ is Heaviside step function. $\rho$ is a parameter to characterize the size of extended source, called extended source parameter, defined by the ratio of the source angle and the Einstein angle: $\rho\equiv \theta_{\rm s}/\theta_{\rm E}$. The real star has the profile darkened toward the limb. One of the most general form of limb-darkening (LD) profile is \citep{Gimenez.Gimenez.2006}
\begin{align}
    s_{\rm LD}(u) &= \frac{1}{N}\left[s_{\rm disk}(u)-\sum_{n=1} u_{\lambda, n}s_n(u)\right] \label{eq:limb-darkening-profile}\\
    s_n(u) &= \frac{1+n/2}{\pi\rho^2}\left(1-\frac{u^2}{\rho^2}\right)^{n/2}\Theta(u-\rho), \label{eq:limb-n-term}
\end{align}
where $u_{\lambda,i}$ are the coefficients depending on the wave length to observe, and $N=1-\sum_nu_{\lambda,n}$ is the normalization. The summation of the above equation is truncated at some order depending on the required accuracy. The first and second terms in the summation, $s_1(u)$ and $s_2(u)$, are widely used in practice, and called as linear limb-darkening and quadratic (or parabolic) limb-darkening \citep{Yoo.Zebrun.2004, Witt.Atrio-Barandela.2019}.

\section{Formalism of FFT based method} \label{sec:fft-based-method}
We first consider the Fourier counterpart of a function which has rotational symmetry, $f(x)=f(\bm{x})$. It also has rotational symmetry and expressed by the zero-th order Hankel transformation of $f(x)$ as
\begin{align}
    \tilde{f}(k) 
    &\equiv \int\diff\bm{x}f(\bm{x})e^{i\bm{x}\cdot\bm{k}} 
    = 2\pi\int_0^\infty \diff x~xf(x) J_0(kx)\\
    &= 2\pi~ \hankel{f(x)}{k}. \label{eq:hankel}
\end{align}
where $J_\nu(x)$ is the $\nu$-th order Bessel function, and ${\cal H}$ is the zero-th order Hankel transformation. The convolution of two functions in configuration space can be expressed by a product of the functions in Fourier space, and hence the Fourier counterpart of Eq.~\eqref{eq:extended-source-magnification} can be expressed as
\begin{align}
    \tilde{A}_{\rm ext}(k) = \tilde{A}_{\rm p}(k)\tilde{s}(k).
    \label{eq:extended-source-magnification-Fourier}
\end{align}
Applying the inverse Fourier transformation or equivalently inverse zero-th order Hankel transformation to Eq.~\eqref{eq:extended-source-magnification-Fourier}, we obtain
\begin{align}
    A_{\rm ext}(u) = \frac{1}{2\pi}~\hankel{\tilde{A}_{\rm p}(k)\tilde{s}(k)}{u}. \label{eq:extended-source-magnification-Hankel}
\end{align}
The Hankel transformation of a given function can be evaluated fast by using FFTLog, a FFT on an equally spaced logarithmic grid \citep{Hamilton.Hamilton.2000}. A single evaluation of the Hankel transformation requires two FFTs (one FFT and one inverse FFT). We have three Hankel transformations in Eq.~\eqref{eq:extended-source-magnification-Hankel}, and hence one evaluation of the extended source magnification includes six FFTs in total. FFT requires computational time of order ${\cal O}(n \log n)$ which is much faster than the two-dimensional integral of order ${\cal O}(n\times n^2)$, where $n$ is the number of points on a grid on $u$ we evaluate the magnification.

There are four points to note. First is that we have to remove a trivial contribution from magnifications for numerical evaluation. Microlensing magnifications always behave $A\rightarrow1$ at $u\rightarrow\infty$, which makes the Fourier counterpart of magnification diverge numerically. In order to get rid of the divergence, we first separate out the contribution as $A_{\rm p}(u)=(A_{\rm p}(u)-1) +1$ and notice that a convolution of the second term and the source profile always gives a trivial contribution: $\int\diff\bm{x}~1\times s(\bm{x}+\bm{u})=1$. Hence, instead of Eq.~\eqref{eq:extended-source-magnification-Hankel} we numerically evaluate 
\begin{align}
    A_{\rm ext}(u)-1 = \frac{1}{2\pi} \hankel{\tilde{A}_{{\rm p}-1}(k)\tilde{s}(k)}{u}. \label{eq:extended-source-magnification-Hankel-without-one}
\end{align}
Here $\tilde{A}_{{\rm p}-1}(k)$ stands for the Hankel transformation of $A_{\rm p}(u)-1$.

The second point is that the numerical evaluation of Eq.~\eqref{eq:extended-source-magnification-Hankel-without-one} by FFT fails when $\rho\rightarrow0$. This can be understood as follows. When $\rho\rightarrow0$, behaviour of the extended source magnification is almost the same as that of the point source magnification, which scales $A_{\rm p}\sim1/u$ at $u\rightarrow0$. Hence, to obtain the extended source magnification, increasing high-$k$ modes need to be included in Eq.~\eqref{eq:extended-source-magnification-Hankel-without-one}, which cannot be covered by a finite range of a FFT grid. However, we can still utilize the FFT expression for the case where $\rho\ll1$ and $u\lesssim \rho$. Since the dominant contribution comes from high-$k$ mode around $k\sim 2\pi/\rho$, we can use an asymptotic form of $A_{{\rm p}-1}(k) \sim 1/k$.
\begin{align}
    A_{\rm ext}(u) -1
    &\sim \frac{1}{2\pi}\int_0^\infty \diff k k \frac{1}{k}\tilde{s}(k)J_0(ku)\\
    &= \frac{1}{\rho} \frac{1}{2\pi}\int_0^\infty\diff x \tilde{s}\left(\frac{x}{\rho}\right)J_0\left(x\frac{u}{\rho}\right)
    \equiv \frac{1}{\rho} A_{{\rm ext}0}(u/\rho). \label{eq:extended-source-magnification-small-rho}
\end{align}
Here, because the source profile function depends on $u$ and $\rho$ through $u/\rho$, the function $A_{{\rm f}0}(x)$ is also a function of $u/\rho$. This enables pre-computation of $A_{{\rm f}0}(x)$ once the source profile type is given without specifying the value of extended source parameter. This formulation does not require any assumption of the source profile type, and hence is applicable to any source profiles. For $\rho\ll1$ and $u\gg\rho$, we can use an asymptotic behavior, $A_{\rm ext}(u)\sim A_{\rm p}(u)$.

The third point to note is that we can obtain analytic expressions for source profiles introduced in Section~\ref{sec:basics-of-microlensing}. The Fourier counterparts of the disk profile and the terms of limb-darkening profile defined in Eq.~\eqref{eq:limb-n-term} are given as
\begin{align}
    &\tilde{s}_{\rm disk}(k) = 2\frac{J_1(k\rho)}{k\rho} \\
    &\tilde{s}_{n}(k) = \frac{\nu 2^{\nu}\Gamma(\nu)}{(k\rho)^{\nu}} J_{\nu}(k\rho),\hspace{2em} \nu=1+\frac{n}{2}
\end{align}
These analytic expressions are useful to reduce the number of FFT calls to compute Eq.~\eqref{eq:extended-source-magnification-Hankel-without-one} and to obtain numerically stable evaluations. 

The last point is about the extended source magnification at $u=0$. Since our method performs FFT on the equally spaced logarithmic grid, it cannot evaluate $A_{\rm ext}(u=0)$. However, $A_{\rm ext}(u=0)$ requires only one evaluation and we can directly evaluate it with low computational cost by
\begin{align}
    A_{\rm ext}(u=0) = \int_0^\rho \diff x~xA(x) s(x).
\end{align}
For the disk and LD profiles, analytic solutions are available, some of which are given by
\begin{align}
    A_{\rm ext}(u=0)=
    \begin{cases}
    \frac{\sqrt{\rho^2+4}}{\rho} & \text{(disk)} \\
    \frac{2(\rho^2+2)E(-\rho^2/4)-(\rho^2+4)K(-\rho^2/4)}{\rho^3} & (\text{linear LD}) \\
    \frac{\rho(\rho^2+2)\sqrt{\rho^2+4}-8{\rm Arcsinh}(\rho/2)}{\rho^4} & \text{(quadratic LD)}
    \end{cases},
\end{align}
where $K(x)$ and $E(x)$ are the complete elliptic integrals of the first and second kind.

In an actual observation, our observable is the total number of photons during a finite exposure time, so the light curve is averaged over the finite exposure time. The magnification with the finite exposure time, $\Delta t$, is given by
\begin{align}
    \bar{A}(t) = \frac{1}{\Delta t}\int_{\Delta t/2}^{-\Delta t/2} \diff t' A(t+t'). \label{eq:exposure-time-magnification}
\end{align}
Applying one-dimensional Fourier transformation to Eq.~\eqref{eq:exposure-time-magnification}, we obtain
\begin{align}
    \bar{A}(t) = \int\frac{\diff f}{2\pi}\tilde{A}(f) \frac{\sin(f\Delta t/2)}{f\Delta t/2}e^{ift},\label{eq:exposure-time-fft}
\end{align}
where Fourier counterpart is given by
\begin{align}
    \tilde{A}(f) = \int\diff t~A(t)e^{-ift}. \label{eq:magnification-freq}
\end{align}
Eqs.~\eqref{eq:exposure-time-fft} and \eqref{eq:magnification-freq} can be evaluated fast by using FFT on a linearly spaced grid.

\section{Result}\label{sec:result}
We implement the FFT based magnification in python using FFTLog developed by \cite{Fang.MacCrann.2020} \footnote{We use a python code available at
\url{https://github.com/xfangcosmo/FFTLog-and-beyond} by Xiao Fang.}. 
Using FFTLog, the Hankel transformation of a given function is performed on an equally spaced logarithmic grid defined by user. We use a FFT grid of $u_i=10^{-6+9i/N}$ where $i=0,\cdots,N-1$ and $N=1024$. We use Eq.~\eqref{eq:extended-source-magnification-Hankel-without-one} for $\rho<10^{-4}$, Eq.~\eqref{eq:extended-source-magnification-small-rho} for $\rho\geq 10^{-4}$ and $u<10\rho$, and Eq.~\eqref{eq:point-source-magnification} for $\rho<10^{-4}$ and $u\geq10\rho$. Once the extended source magnification is evaluated on the FFT grid, we interpolate it to obtain the magnification on the desired $u$.

\begin{figure*}
    \centering
    \includegraphics[width=0.5\textwidth]{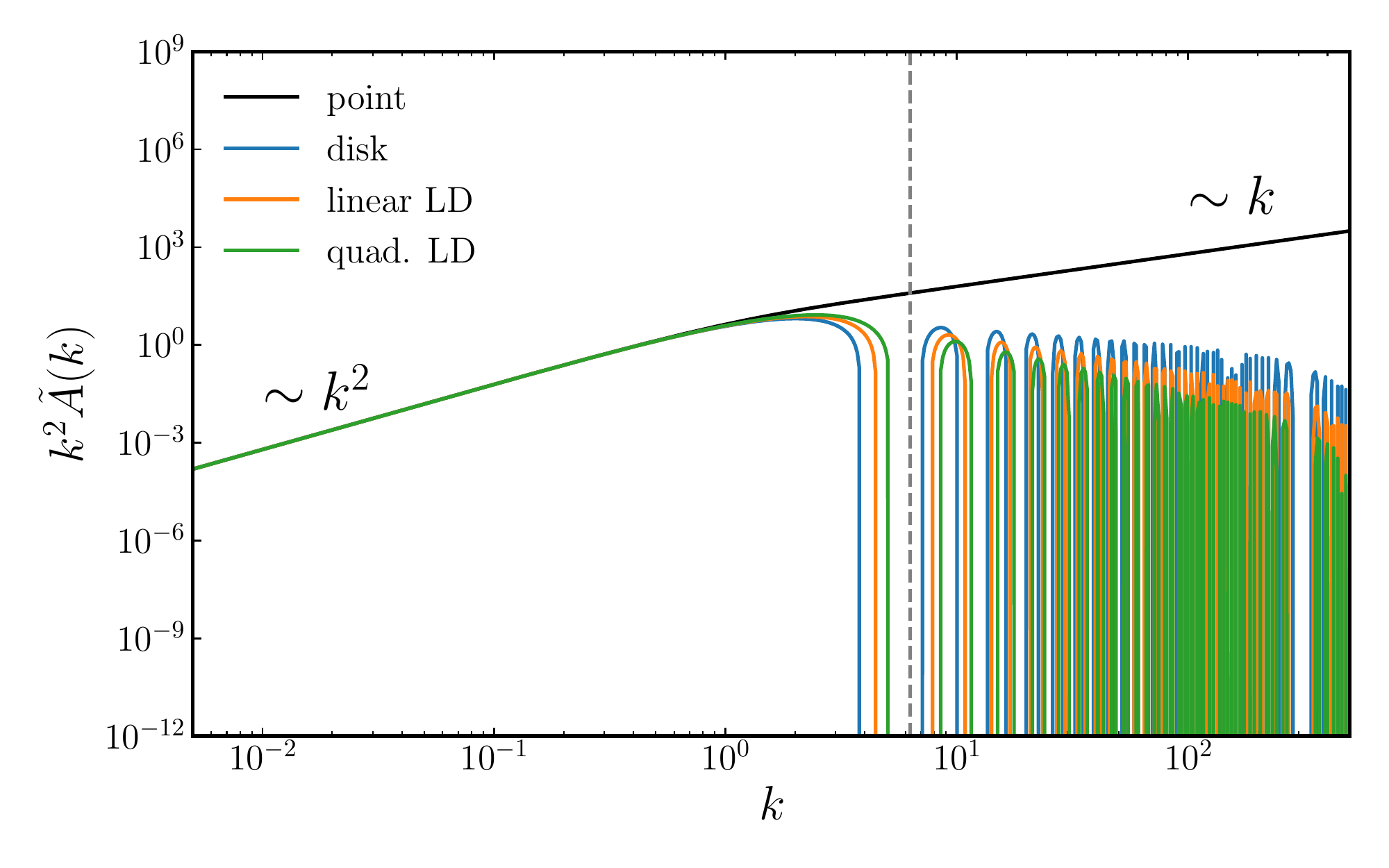}
    \caption{Fourier counterparts of magnifications. Black line shows Fourier counterpart of point source magnification, $k^2 \tilde{A}_{{\rm p}-1}(k)$, which has asymptotic form of $\sim k^2$ at $k\ll1$ and $\sim k$ at $k\gg1$. The colored lines show the Fourier counterparts of the extended source magnifications for different source profile types, $k^2\tilde{A}_{{\rm p}-1}(k)\tilde{s}(k)$, for disk, linear LD, and parabolic/quadratic LD profiles. The extended source parameter is fixed, $\rho=1$, for all the source profiles. The vertical dashed line shows the typical source size scale $k_{\rho}=2\pi/\rho$, beyond which the Fourier modes are dumped reflecting flattened magnification by extended source effect. Note that extended source Fourier counterparts include additional dumping factor of  $\exp[-(k/50k_\rho)^2]$ to filter out the sparsely sampled (noisy) high-$k$ modes.}
    \label{fig:ak-rho1}
\end{figure*}

Fig.~\ref{fig:ak-rho1} shows the Fourier counterparts of magnifications. The extended source Fourier counterparts with disk, linear limb-darkening and quadratic limb-darkening are shown, and the extended source parameter is fixed, $\rho=1$. The Fourier counterpart of the point source magnification has the asymptotic form of $k^2\tilde{A}_{{\rm p}-1}\sim k^2$ at $k\ll1$ and $\sim k$ at $k\gg1$. The former asymptotic behaviour is derived as 
\begin{align}
    k^2\tilde{A}_{{\rm p}-1}(k) = 2\pi \int \diff x~xJ_0(x)[A_{\rm p}(x/k)-1] \sim 2\pi\int\diff x~xJ_0(x) \frac{2k^2}{x\sqrt{x^2+4k^2}} \propto k^2 \hspace{2em}(k\ll1).
\end{align}
The latter asymptotic behaviour is due to the asymptotic behaviour of $A_{\rm p}(u)-1 \sim u^{-1}$ at $u\ll1$, which enables the evaluation for small $\rho$ as described around Eq.~\eqref{eq:extended-source-magnification-small-rho}. The Fourier counterpart of the extended source magnification is given by a product of those of the point source magnification and the source profile, $k^2\tilde{A}_{{\rm p}-1}(k)\tilde{s}(k)$. Since the extended source effect flattens the point source magnifications within a typical radius $u\lesssim\rho$, high-$k$ modes of the extended source Fourier counterpart is dumped beyond the corresponding scale $k_{\rho}\sim 2\pi/\rho$. As $\rho$ becomes smaller, $k_{\rho}$ becomes larger and more high-$k$ modes contribute and the extended source magnification approaches to the point source magnification. From the normalization condition of source profile, $\tilde{s}(k=0)=2\pi\int \diff u~us(u)=1$ holds, and hence the Fourier counterparts of the extended source magnifications approaches to that of point source magnification for any profile at $k\rightarrow0$. Since high-$k$ modes are noisy due to the sparse sampling on FFT grid, we multiply a exponential dumping factor, $\exp[-(k/50k_\rho)^2]$, to the Fourier counterpart of the extended source magnification.

\begin{figure}
    \centering
    \includegraphics[width=1\textwidth]{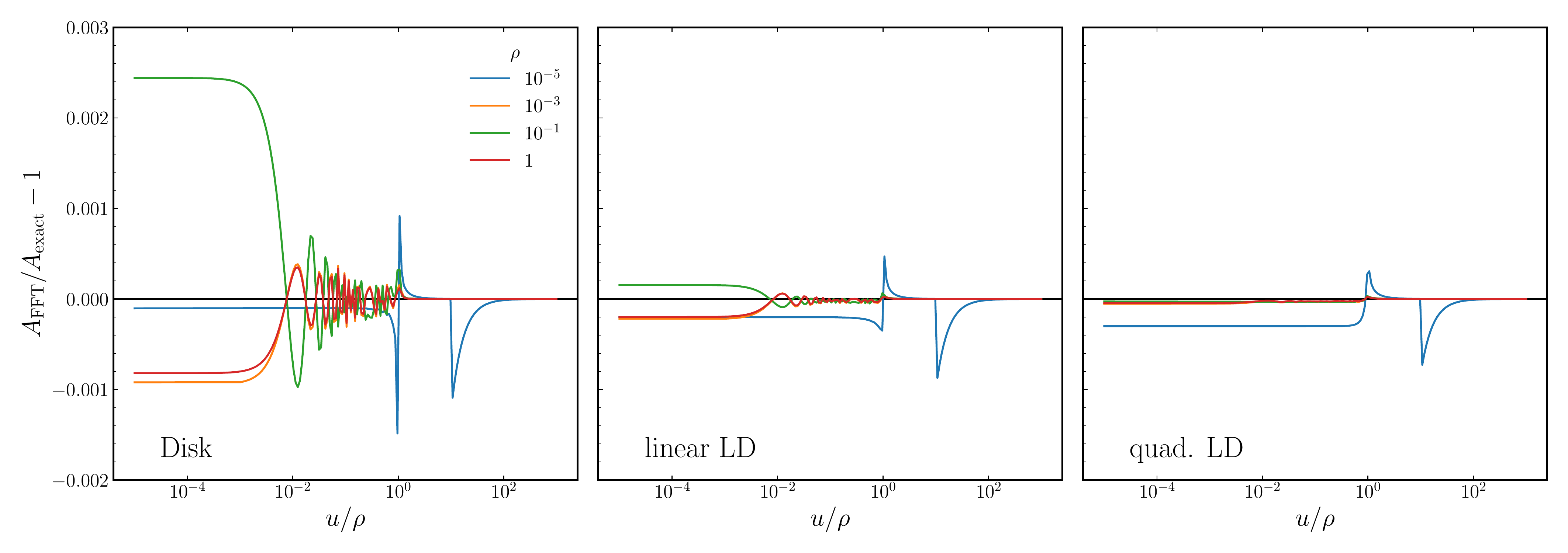}
    \caption{The residuals of magnification evaluation by the FFT based method. The reference magnification is the direct evaluation of Eq.~\eqref{eq:extended-source-magnification}. From the left to right panels, results for disk, linear LD and quadratic LD profiles are shown. The different colors shows the different extended source parameters as indicated in the legend on the leftmost panel. The x-axes are normalized by the extended source parameter $\rho$.}
    \label{fig:afft-residual}
\end{figure}

\begin{figure}
    \centering
    \includegraphics[width=1\textwidth]{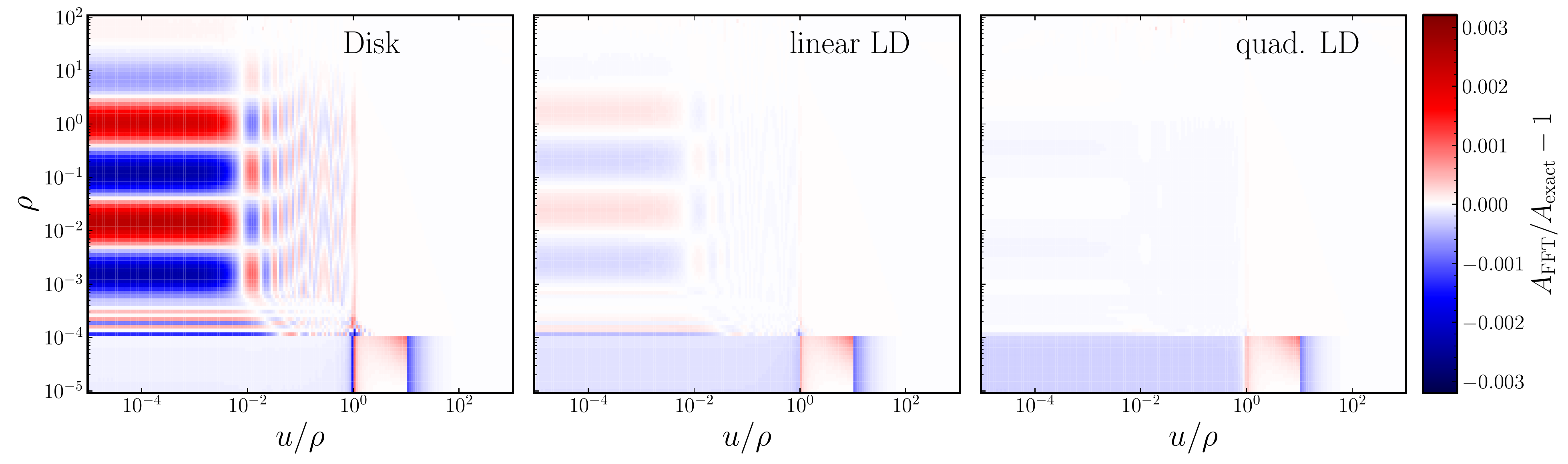}
    \caption{The residuals of magnification evaluation by the FFT based method in two parameter space, $(u/\rho, \rho)$. The reference magnification is the direct evaluation of Eq.~\eqref{eq:extended-source-magnification}. The different panels shows for the different profiles. The x-axes are normalized by $\rho$. In the region of $\rho<10^{-4}$, the behaviour of residual is different from $\rho>10^{-4}$, because we use Eqs.~\eqref{eq:extended-source-magnification-small-rho} and \eqref{eq:point-source-magnification} for $u<10\rho$ and $u\geq10\rho$ respectively.}
    \label{fig:afft-residual-2d}
\end{figure}

Fig.~\ref{fig:afft-residual} shows the residual of the evaluation by the FFT based method developed in this paper, for the different source sizes and for the different source profiles. The reference magnification is evaluated by directly performing the two-dimensional integration of Eq.~\ref{eq:extended-source-magnification}, using the integration routine {\tt scipy.integrate.quad} twice. Since the extended source effect is significant in $u\lesssim\rho$ region, we normalize $u$ with $\rho$ in the x-axes. We can see that the FFT based method achieves accuracy better than 0.3\% over the plot scales for every source size and every profile. The implementation for small $\rho$ described around Eq.~\eqref{eq:extended-source-magnification-small-rho} also achieves 0.2\% accuracy, as shown by result for $\rho=10^{-5}$. Note that the accuracy is better for higher order of LD profile. This is because the higher order profile decreases faster at $k>k_\rho$ as shown in Fig.~\ref{fig:ak-rho1}, and high-$k$ modes contribute relatively less. Fig.~\ref{fig:afft-residual-2d} shows more comprehensive plot of residual in the two parameter space, $(u/\rho, \rho)$.

\begin{figure}
    \centering
    \includegraphics[width=1\textwidth]{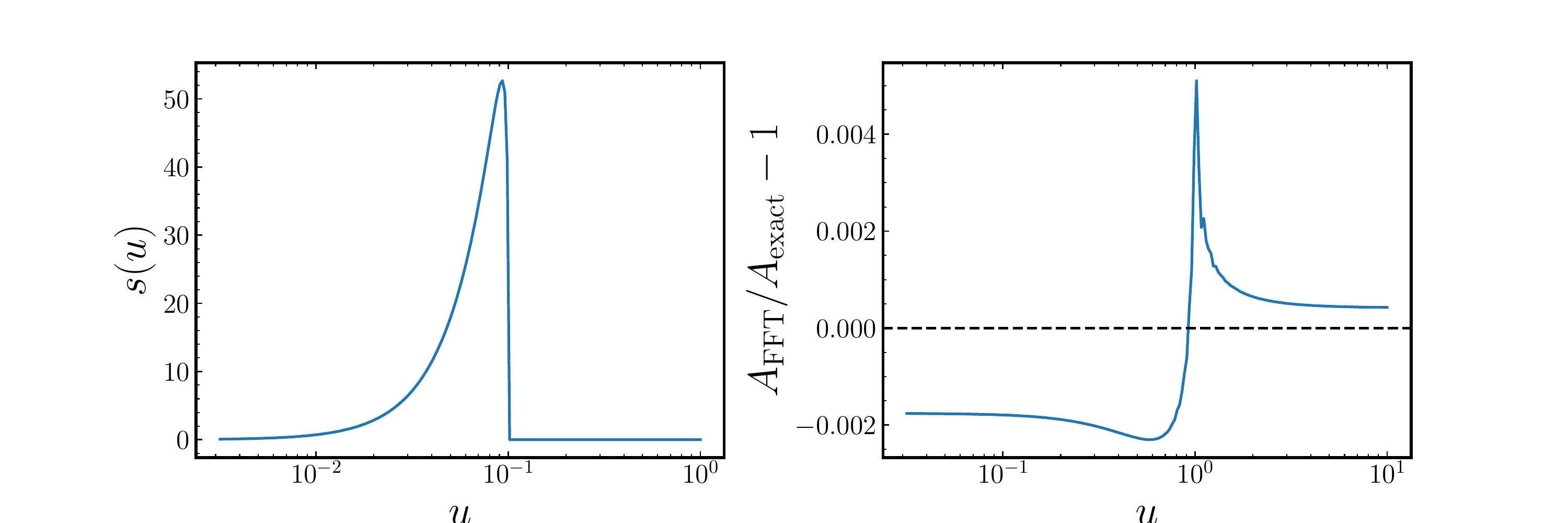}
    \caption{A demonstration of the FFT based method for a source profile for which analytic expression of the Fourier counterpart is unavailable. The left panel shows the logarithmic source profile with $\rho=0.1$, and the right panel shows the residual of the FFT based method.}
    \label{fig:log-profile}
\end{figure}

Up to now, we only focus on disk, linear limb-darkening and quadratic limb-darkening profiles, for which analytic expressions of the Fourier counterparts are available. In Fig.~\ref{fig:log-profile}, we demonstrate that the FFT based method is applicable even when the analytic expression of the Fourier counterpart for the extended source profile is unavailable. We take as an example the logarithmic profile proposed by \cite{Klinglesmith.Sobieski.1970}, $s(u) \propto \sqrt{1-u^2/\rho^2}\log\sqrt{1-u^2/\rho^2}$, which fits early-type star profile on the top of disk and linear limb-darkening profile \footnote{To be precise, an analytic expression of the Fourier counterpart is available for the logarithmic profile. It, however, includes derivative of Bessel function with respect to order, which requires additional caution for the numerical evaluation, and hence we take it as an example of cases for which the analytic expression of the Fourier counterpart is unavailable.}. We numerically evaluate the Fourier counterpart of the logarithmic source profile with Eq.~\eqref{eq:hankel} by FFTLog. Fig.~\ref{fig:log-profile} shows that the FFT based method works with accuracy better than 0.5\%, meaning that the FFT based is useful even when the analytic expression of the Fourier counterpart for the source profile is unavailable, e.g. the source profile calibrated by simulation \citep{Orosz.Hauschildt.2000}.

\begin{figure}
    \centering
    \includegraphics[width=0.7\textwidth]{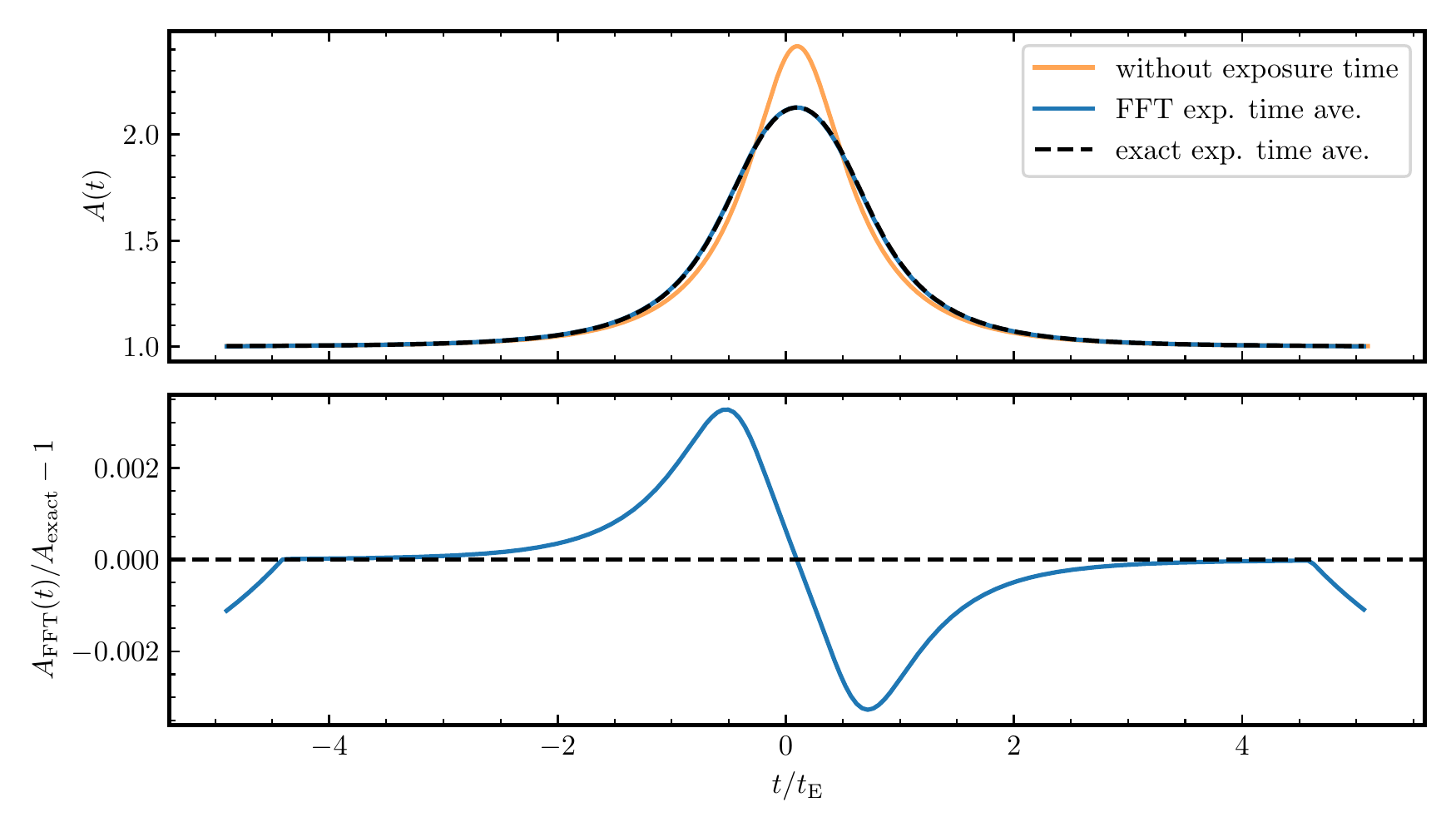}
    \caption{The performance of the FFT based evaluation of the exposure time averaged magnification. Here, lensing parameters are $(t_0, t_{\rm E}, u_{\rm min}, \rho, \Delta t)=(10~{\rm sec}, 100~{\rm sec}, 0.1, 0.1, 100~{\rm sec})$. The exact reference magnification is evaluated by {\tt scipy.integrate.quad} function. The time on x-axis is normalized by the Einstein time, $t_{\rm E}$.}
    \label{fig:afft-timeave}
\end{figure}

In Fig.~\ref{fig:afft-timeave}, we show the performance of the FFT based evaluation of the exposure time averaged magnification discussed in Section~\ref{sec:fft-based-method}. Here we use $n=2048$ grid on time. As in the previous two sections, exact reference magnification is evaluated by {\tt scipy.integrate.quad} function. With our choice of time grid, the FFT based method achieves accuracy better than 0.3\%.

\begin{figure}
    \centering
    \includegraphics[width=1\textwidth]{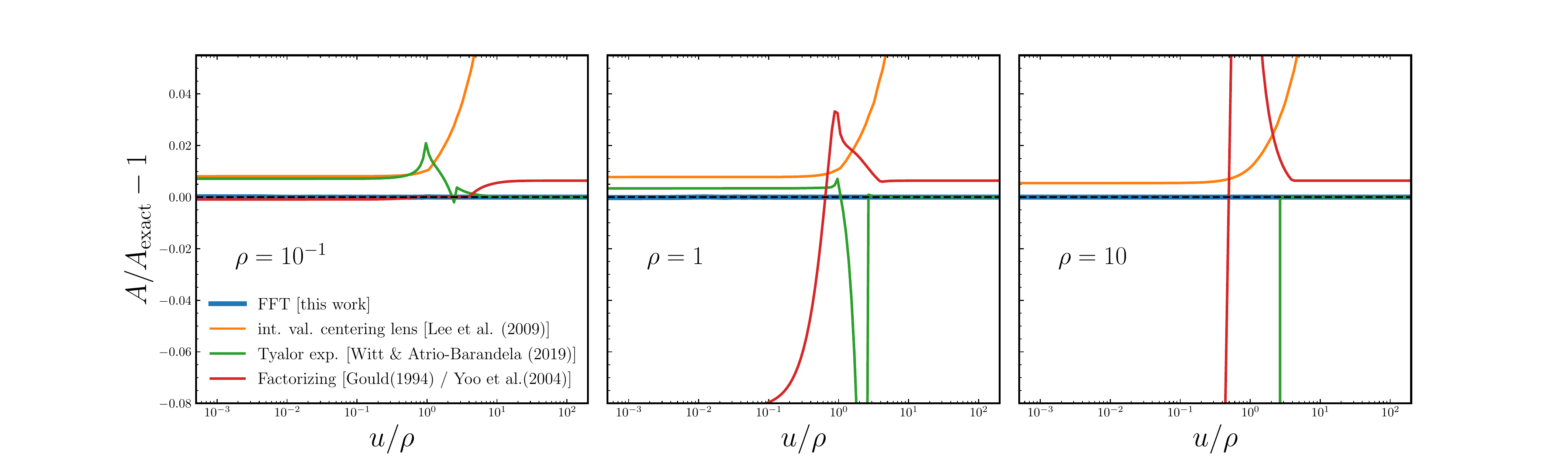}
    \caption{Comparison of accuracies with other studies. Different panels show the comparison of precision for different extended source parameters, $\rho=10^{-1}$, $1$ and $10$. Profile type is fixed to linear limb-darkening. The lines with different colors show methods of this paper and other studies (see the main text for detail). X-axes are normalized by $\rho$.}
    \label{fig:comparison}
\end{figure}

Fig.~\ref{fig:comparison} shows comparison of accuracies between methods developed in this paper and other studies for the different extended source parameter, $\rho$. We only compare the results for the linear limb-darkening profile, for which all the methods are developed. The method by \cite{Lee.Bender.2009} computes the convolution in Eq.~\eqref{eq:extended-source-magnification} with integration variables on polar coordinate centering the lens object to avoid numerical instability of Eq.~\eqref{eq:point-source-magnification} at $u=0$. The method includes the two-dimensional integral for the general profile and can be reduced to one-dimensional integral only for disk profile. Here we implement the method with $200\times200$ grid on the polar coordinate, $(\tilde{u}, \theta)$ (see Eq.~(13) of \cite{Lee.Bender.2009} for detail). \cite{Witt.Atrio-Barandela.2019} derived analytic approximations using the Taylor expansion of Eq.~\eqref{eq:point-source-magnification}. The accuracy of the method becomes worse when $u, \rho \gtrsim1$ because of the Taylor expansion, and hence they uses the analytic result of the extended source magnification with disk profile \citep{Witt.Mao.1994} at $u>3\rho$. \cite{Gould.Gould.1994} derived a method which factorizes out the extended source effect for disk profile. \cite{Yoo.Zebrun.2004} generalized the method to compute for the linear limb-darkening profile. From the comparison plot, we can see that the FFT based method developed in this paper works an order of magnitude better than other methods.

As described in the beginning of this section, the FFT based method evaluates the extended source magnification on a fixed FFT grid and interpolate it to desired $u$ points. Hence the computational time does not scales with the number of the desired $u$ points. The FFT based method developed in this paper takes $\sim 1.4$ msec (CPU time on a laptop) to compute the extended source magnification for liner limb-darkening profile. This computational time can be compared to those of other methods\footnote{I implemented the methods by other studies in python except for method by \cite{Lee.Bender.2009}. I implemented the method by \cite{Lee.Bender.2009} in C because it includes two for loops over the two integration variables and slow if implemented in python.}: $\sim 830$ msec with method by \cite{Lee.Bender.2009}, $\sim 93$ msec with method by \cite{Witt.Atrio-Barandela.2019}, and $\sim 2.8$ msec with method by \cite{Gould.Gould.1994} and \cite{Yoo.Zebrun.2004}. We also note that the FFT based method for the exposure time averaged magnification is also as fast as $\sim 1.0$ msec.

\section{Conclusion}
We developed a new method which evaluates the microlensing magnification with the extended source and the exposure time fast and accurately based on FFT. The FFT based method performs the Hankel transformations to compute convolution of the point source magnification and the source profile, and performs usual FFT to compute the exposure time averaged magnification. Our method achieves 0.3\% accuracy everywhere in the parameter space, $(u, \rho)$, which is much better than the methods developed by the other studies. The FFT based method takes only $3$ msec, which is as fast as the method by \cite{Gould.Gould.1994, Yoo.Zebrun.2004} and faster than any other studies. The FFT based method is applicable even when analytic expression of the Fourier counterpart for the source profile is unavailable, and hence easy to evaluate the magnification with any source profiles. Codes are available on github repository at \url{https://github.com/git-sunao/fft-extended-source}.

\begin{acknowledgments}
I thanks Toshiki Kurita for useful discussion to check the formalism, Kento Masuda for the references of source profiles, Xiao Fang for allowing to distribute my code with his FFTLog code, and Masahiro Takada for reviewing the draft. This work is supported by JSPS KAKENHI Grant Number 21J10314. This work is supported by World Premier International Research Center Initiative (WPI Initiative), MEXT, Japan. This work is also supported by International Graduate Program for Excellence in Earth-Space Science (IGPEES), World-leading Innovative Graduate Study (WINGS) Program, the University of Tokyo.
\end{acknowledgments}
\vspace{5mm}

\bibliography{refs}{}

\begin{thebibliography}{}
\expandafter\ifx\csname natexlab\endcsname\relax\def\natexlab#1{#1}\fi
\providecommand{\url}[1]{\href{#1}{#1}}
\providecommand{\dodoi}[1]{doi:~\href{http://doi.org/#1}{\nolinkurl{#1}}}
\providecommand{\doeprint}[1]{\href{http://ascl.net/#1}{\nolinkurl{http://ascl.net/#1}}}
\providecommand{\doarXiv}[1]{\href{https://arxiv.org/abs/#1}{\nolinkurl{https://arxiv.org/abs/#1}}}

\bibitem[{{Alard} {et~al.}(1995){Alard}, {Guibert}, {Bienayme}, {Valls-Gabaud},
  {Robin}, {Terzan}, \& {Bertin}}]{Alard.Bertin.1995}
{Alard}, C., {Guibert}, J., {Bienayme}, O., {et~al.} 1995, The Messenger, 80,
  31

\bibitem[{Alcock {et~al.}(1993)Alcock, Akerlof, Allsman, Axelrod, Bennett,
  Chan, Cook, Freeman, Griest, Marshall, Park, Perlmutter, Peterson, Pratt,
  Quinn, Rodgers, Stubbs, \& Sutherland}]{Alcock.Sutherland.1993}
Alcock, C., Akerlof, C.~W., Allsman, R.~A., {et~al.} 1993, Nature, 365, 621,
  \dodoi{10.1038/365621a0}

\bibitem[{Aubourg {et~al.}(1993)Aubourg, Bareyre, Bréhin, Gros, Lachièze-Rey,
  Laurent, Lesquoy, Magneville, Milsztajn, Moscoso, Queinnec, Rich, Spiro,
  Vigroux, Zylberajch, Ansari, Cavalier, Moniez, Beaulieu, Ferlet, Grison,
  Vidal-Madjar, Guibert, Moreau, Tajahmady, Maurice, Prévôt, \&
  Gry}]{Aubourg.Gry.1993}
Aubourg, E., Bareyre, P., Bréhin, S., {et~al.} 1993, Nature, 365, 623,
  \dodoi{10.1038/365623a0}

\bibitem[{Bond {et~al.}(2001)Bond, Abe, Dodd, Hearnshaw, Honda, Jugaku,
  Kilmartin, Marles, Masuda, Matsubara, Muraki, Nakamura, Nankivell, Noda,
  Noguchi, Ohnishi, Rattenbury, Reid, Saito, Sato, Sekiguchi, Skuljan,
  Sullivan, Sumi, Takeuti, Watase, Wilkinson, Yamada, Yanagisawa, \&
  Yock}]{Bond.Yock.2001}
Bond, I., Abe, F., Dodd, R., {et~al.} 2001, Monthly Notices of the Royal
  Astronomical Society, 327, 868, \dodoi{10.1046/j.1365-8711.2001.04776.x}

\bibitem[{{Einstein}(1916)}]{Einstein.Einstein.1916}
{Einstein}, A. 1916, Annalen der Physik, 354, 769,
  \dodoi{10.1002/andp.19163540702}

\bibitem[{Fang {et~al.}(2020)Fang, Krause, Eifler, \&
  MacCrann}]{Fang.MacCrann.2020}
Fang, X., Krause, E., Eifler, T., \& MacCrann, N. 2020, Journal of Cosmology
  and Astroparticle Physics, 2020, 010, \dodoi{10.1088/1475-7516/2020/05/010}

\bibitem[{Gaudi {et~al.}(2019)Gaudi, Akeson, Anderson, Bachelet, Bennett,
  Bhattacharya, Bozza, Novati, Henderson, Johnson, Kruk, Lu, Mao, Montet,
  Nataf, Penny, Poleski, Ranc, Sahu, Shvartzvald, Spergel, Suzuki, Stassun, \&
  Street}]{Gaudi.Street.2019}
Gaudi, B.~S., Akeson, R., Anderson, J., {et~al.} 2019, arXiv

\bibitem[{Giménez(2006)}]{Gimenez.Gimenez.2006}
Giménez, A. 2006, {Equations for the analysis of the light curves of
  extra-solar planetary transits}.
\newblock \url{https://www.aanda.org/articles/aa/pdf/2006/18/aa4445-05.pdf}

\bibitem[{Gould(1994)}]{Gould.Gould.1994}
Gould, A. 1994, The Astrophysical Journal, 421, L71, \dodoi{10.1086/187190}

\bibitem[{Griest {et~al.}(2014)Griest, Cieplak, \& Lehner}]{Griest.Lehner.2014}
Griest, K., Cieplak, A.~M., \& Lehner, M.~J. 2014, The Astrophysical Journal,
  786, 158, \dodoi{10.1088/0004-637x/786/2/158}

\bibitem[{Hamilton(2000)}]{Hamilton.Hamilton.2000}
Hamilton, A. J.~S. 2000, Monthly Notices of the Royal Astronomical Society,
  312, 257, \dodoi{10.1046/j.1365-8711.2000.03071.x}

\bibitem[{Ivezi\'c {et~al.}(2019)Ivezi\'c, Kahn, Tyson, Abel, Acosta, Allsman,
  Alonso, AlSayyad, Anderson, Andrew, Angel, Angeli, Ansari, Antilogus, Araujo,
  Armstrong, Arndt, Astier, Aubourg, Auza, Axelrod, Bard, Barr, Barrau,
  Bartlett, Bauer, Bauman, Baumont, Bechtol, Bechtol, Becker, Becla, Beldica,
  Bellavia, Bianco, Biswas, Blanc, Blazek, Blandford, Bloom, Bogart, Bond,
  Booth, Borgland, Borne, Bosch, Boutigny, Brackett, Bradshaw, Brandt, Brown,
  Bullock, Burchat, Burke, Cagnoli, Calabrese, Callahan, Callen, Carlin,
  Carlson, Chandrasekharan, Charles-Emerson, Chesley, Cheu, Chiang, Chiang,
  Chirino, Chow, Ciardi, Claver, Cohen-Tanugi, Cockrum, Coles, Connolly, Cook,
  Cooray, Covey, Cribbs, Cui, Cutri, Daly, Daniel, Daruich, Daubard, Daues,
  Dawson, Delgado, Dellapenna, Peyster, Val-Borro, Digel, Doherty, Dubois,
  Dubois-Felsmann, Durech, Economou, Eifler, Eracleous, Emmons, Neto, Ferguson,
  Figueroa, Fisher-Levine, Focke, Foss, Frank, Freemon, Gangler, Gawiser,
  Geary, Gee, Geha, Gessner, Gibson, Gilmore, Glanzman, Glick, Goldina,
  Goldstein, Goodenow, Graham, Gressler, Gris, Guy, Guyonnet, Haller, Harris,
  Hascall, Haupt, Hernandez, Herrmann, Hileman, Hoblitt, Hodgson, Hogan,
  Howard, Huang, Huffer, Ingraham, Innes, Jacoby, Jain, Jammes, Jee, Jenness,
  Jernigan, Jevremović, Johns, Johnson, Johnson, Jones, Juramy-Gilles, Jurić,
  Kalirai, Kallivayalil, Kalmbach, Kantor, Karst, Kasliwal, Kelly, Kessler,
  Kinnison, Kirkby, Knox, Kotov, Krabbendam, Krughoff, Kubánek, Kuczewski,
  Kulkarni, Ku, Kurita, Lage, Lambert, Lange, Langton, Guillou, Levine, Liang,
  Lim, Lintott, Long, Lopez, Lotz, Lupton, Lust, MacArthur, Mahabal,
  Mandelbaum, Markiewicz, Marsh, Marshall, Marshall, May, McKercher, McQueen,
  Meyers, Migliore, Miller, Mills, Miraval, Moeyens, Moolekamp, Monet, Moniez,
  Monkewitz, Montgomery, Morrison, Mueller, Muller, Arancibia, Neill, Newbry,
  Nief, Nomerotski, Nordby, O’Connor, Oliver, Olivier, Olsen, O’Mullane,
  Ortiz, Osier, Owen, Pain, Palecek, Parejko, Parsons, Pease, Peterson,
  Peterson, Petravick, Petrick, Petry, Pierfederici, Pietrowicz, Pike, Pinto,
  Plante, Plate, Plutchak, Price, Prouza, Radeka, Rajagopal, Rasmussen,
  Regnault, Reil, Reiss, Reuter, Ridgway, Riot, Ritz, Robinson, Roby, Roodman,
  Rosing, Roucelle, Rumore, Russo, Saha, Sassolas, Schalk, Schellart,
  Schindler, Schmidt, Schneider, Schneider, Schoening, Schumacher, Schwamb,
  Sebag, Selvy, Sembroski, Seppala, Serio, Serrano, Shaw, Shipsey, Sick,
  Silvestri, Slater, Smith, Smith, Sobhani, Soldahl, Storrie-Lombardi, Stover,
  Strauss, Street, Stubbs, Sullivan, Sweeney, Swinbank, Szalay, Takacs, Tether,
  Thaler, Thayer, Thomas, Thornton, Thukral, Tice, Trilling, Turri, Berg, Berk,
  Vetter, Virieux, Vucina, Wahl, Walkowicz, Walsh, Walter, Wang, Wang, Warner,
  Wiecha, Willman, Winters, Wittman, Wolff, Wood-Vasey, Wu, Xin, Yoachim, \&
  Zhan}]{Ivezic.Zhan.2019}
Ivezi\'c, v., Kahn, S.~M., Tyson, J.~A., {et~al.} 2019, The Astrophysical
  Journal, 873, 111, \dodoi{10.3847/1538-4357/ab042c}

\bibitem[{Kim {et~al.}(2010)Kim, Park, Lee, Yuk, Han, O'Brien, Gould, Lee, \&
  Kim}]{Kim.Kim.2010}
Kim, S.-L., Park, B.-G., Lee, C.-U., {et~al.} 2010, Ground-based and Airborne
  Telescopes III, 77333F, \dodoi{10.1117/12.856833}

\bibitem[{Klinglesmith \& Sobieski(1970)}]{Klinglesmith.Sobieski.1970}
Klinglesmith, D.~A., \& Sobieski, S. 1970, The Astronomical Journal, 75, 175,
  \dodoi{10.1086/110960}

\bibitem[{Laureijs {et~al.}(2012)Laureijs, Gondoin, Duvet, Criado, Hoar,
  Amiaux, Auguères, Cole, Cropper, Ealet, Ferruit, Sanz, Jahnke, Kohley,
  Maciaszek, Mellier, Oosterbroek, Pasian, Sauvage, Scaramella, Sirianni, \&
  Valenziano}]{Laureijs.Valenziano.2012}
Laureijs, R., Gondoin, P., Duvet, L., {et~al.} 2012, Space Telescopes and
  Instrumentation 2012: Optical, Infrared, and Millimeter Wave, 84420T,
  \dodoi{10.1117/12.926496}

\bibitem[{Lee {et~al.}(2009)Lee, Riffeser, Seitz, \& Bender}]{Lee.Bender.2009}
Lee, C.-H., Riffeser, A., Seitz, S., \& Bender, R. 2009, The Astrophysical
  Journal, 695, 200, \dodoi{10.1088/0004-637x/695/1/200}

\bibitem[{Niikura {et~al.}(2019)Niikura, Takada, Yasuda, Lupton, Sumi, More,
  Kurita, Sugiyama, More, Oguri, \& Chiba}]{Niikura.Chiba.2019}
Niikura, H., Takada, M., Yasuda, N., {et~al.} 2019, Nature Astronomy, 3, 524,
  \dodoi{10.1038/s41550-019-0723-1}

\bibitem[{Orosz \& Hauschildt(2000)}]{Orosz.Hauschildt.2000}
Orosz, J.~A., \& Hauschildt, P.~H. 2000, arXiv

\bibitem[{Paczynski(1986)}]{Paczynski.Paczynski.1986}
Paczynski, B. 1986, The Astrophysical Journal, 304, 1, \dodoi{10.1086/164140}

\bibitem[{Refsdal(1964)}]{Refsdal.Bondi.1964}
Refsdal, S. 1964, Monthly Notices of the Royal Astronomical Society, 128, 295,
  \dodoi{10.1093/mnras/128.4.295}

\bibitem[{Shvartzvald \& Maoz(2012)}]{Shvartzvald.Maoz.2012}
Shvartzvald, Y., \& Maoz, D. 2012, Monthly Notices of the Royal Astronomical
  Society, 419, 3631, \dodoi{10.1111/j.1365-2966.2011.20014.x}

\bibitem[{Spergel {et~al.}(2015)Spergel, Gehrels, Baltay, Bennett,
  Breckinridge, Donahue, Dressler, Gaudi, Greene, Guyon, Hirata, Kalirai,
  Kasdin, Macintosh, Moos, Perlmutter, Postman, Rauscher, Rhodes, Wang,
  Weinberg, Benford, Hudson, Jeong, Mellier, Traub, Yamada, Capak, Colbert,
  Masters, Penny, Savransky, Stern, Zimmerman, Barry, Bartusek, Carpenter,
  Cheng, Content, Dekens, Demers, Grady, Jackson, Kuan, Kruk, Melton, Nemati,
  Parvin, Poberezhskiy, Peddie, Ruffa, Wallace, Whipple, Wollack, \&
  Zhao}]{Spergel.Zhao.2015}
Spergel, D., Gehrels, N., Baltay, C., {et~al.} 2015, arXiv

\bibitem[{Udalski {et~al.}(2015)Udalski, Szymański, \&
  Szymański}]{Udalski.Szymanski.2015}
Udalski, A., Szymański, M.~K., \& Szymański, G. 2015, arXiv

\bibitem[{Witt \& Atrio-Barandela(2019)}]{Witt.Atrio-Barandela.2019}
Witt, H.~J., \& Atrio-Barandela, F. 2019, The Astrophysical Journal, 880, 152,
  \dodoi{10.3847/1538-4357/ab2a04}

\bibitem[{Witt \& Mao(1994)}]{Witt.Mao.1994}
Witt, H.~J., \& Mao, S. 1994, The Astrophysical Journal, 430, 505,
  \dodoi{10.1086/174426}

\bibitem[{Yoo {et~al.}(2004)Yoo, DePoy, Gal‐Yam, Gaudi, Gould, Han, Lipkin,
  Maoz, Ofek, Park, Pogge, Udalski, Soszyński, Wyrzykowski, Kubiak,
  Szymański, Pietrzyński, Szewczyk, \& Żebruń}]{Yoo.Zebrun.2004}
Yoo, J., DePoy, D.~L., Gal‐Yam, A., {et~al.} 2004, The Astrophysical Journal,
  603, 139, \dodoi{10.1086/381241}

\end{thebibliography}
\bibliographystyle{aasjournal}



\end{document}